\documentclass[aps,pre,twocolumn,groupedaddress,showpacs,floatfix]{revtex4}
\usepackage{graphicx}
\usepackage{dcolumn}
\usepackage{bm}
\input epsf
\epsfclipon
\begin{document}
\title{Networks with given two-point correlations: hidden correlations from degree correlations}
\author{Agata Fronczak and Piotr Fronczak}
\affiliation{Faculty of Physics and Center of Excellence for
Complex Systems Research, Warsaw University of Technology,
Koszykowa 75, PL-00-662 Warsaw, Poland}
\date{\today}

\begin{abstract}
The paper orders certain important issues related to both
uncorrelated and correlated networks with hidden variables. In
particular, we show that networks being uncorrelated at the hidden
level are also lacking in correlations between node degrees. The
observation supported by the depoissonization idea allows to
extract distribution of hidden variables from a given node degree
distribution. It completes the algorithm for generating
uncorrelated networks that was suggested by other authors. In this
paper we also carefully analyze the interplay between hidden
attributes and node degrees. We show how to extract hidden
correlations from degree correlations. Our derivations provide
mathematical background for the algorithm for generating
correlated networks that was proposed by Bogu\~{n}\'{a} and
Pastor-Satorras.
\end{abstract}


\maketitle

\section{Introduction}\label{intro}


\par Recently, the techniques of equilibrium and non-equilibrium
statistical physics were developed to study complex networks
\cite{0a,0b,BArev}. This paper is devoted to equilibrium
correlated networks, with a special attention given to networks
with two point correlations \cite{newPRL2002,newPRE2003a,
dorCM2002,bogPRE2003,dorCM2003}.

\par What does it mean that a network is correlated? In simple words
one can say that there exist certain relationships between network
nodes. For example, when one considers a social network i.e. a
group of people with links given by acquaintance ties one may
expect that young people are mostly surrounded by other young
people. One may also expect that wealthy individuals are more
often associated with other wealthy individuals than with poor
once. In some sense, the above examples let one suppose that
social networks are positively correlated, at least when one
considers individual's age or income. The situation is more
contentious if one asks for relationship between gender of the
nearest neighbors. Now, it is difficult to assess if social
networks are positively or negatively correlated. The above
examples show that even taking into account a single network one
can observe correlations at different levels of the system
complexity. Each node $i$ in such a system has assigned a set of
different attributes like: gender $g_{i}$, age $a_{i}$, education
$e_{i}$, {\it attractiveness} $k_{i}$ etc. The last property may
be quantified as a number of nearest neighbors of the considered
individual. In the graph theory \cite{wilson_book} the quantity is
known as node's degree.

\par The above outlined network correlations and multilevel structure
constitute two main issues undertaken in this paper.

\par Multilevel topology has been recently considered by several groups
of researchers \cite{gohPRL2001,calPRL2002,sodPRE2002,sodPRE2003a,
sodPRE2003b} and, at present, the proposed models are known as
random networks with hidden variables. To be precise, to our
knowledge, none of the proposed models considers the number of
complexity levels larger than two \cite{bogPRE2003} i.e. each node
$i$ is characterized by at most two parameters: hidden attribute
$h_{i}$ and node's degree $k_{i}$. In general, random networks
with hidden variables have a fixed number of vertices $N$. Each
node in a network belonging to this class of models has assigned a
hidden variable $h_{i}$ (fitness, tag) randomly drawn from a fixed
probability distribution $R(h)$ (throughout the paper we use the
symbol $R$ with reference to distributions at hidden level and $P$
with reference to node degrees). Edges are assigned to pairs of
vertices $\{i,j\}$ with a given connection probability $r_{ij}$.
In the simplest case $r_{ij}$ depends only on values of the hidden
variables $h_{i}$ and $h_{j}$, but in a more general situation it
may for example involve hidden variables characterizing the
nearest or the next-nearest neighbors of both considered nodes $i$
and $j$. In fact, the first case represents networks with at most
two-point correlations at the level of hidden variables, while the
latter one allows for higher order correlations.

\par Below we outline the concept of network
correlations in a more rigorous way. The introduced ideas will be
completed and widely discussed in the next section.

\par From the mathematical point of view, the lack of two point
correlations means that the probability $R(h_{j}/h_{i})$, that an
edge departing from a vertex $i$ of property $h_{i}$ arrives at a
vertex $j$ of property $h_{j}$, is independent of the initial
vertex $i$ \cite{newPRE2001,afCM2005a}. The above translates into
the fact that the nearest neighborhood of each node is the same
(in statistical terms). On the other hand, when $R(h_{j}/h_{i})$
depends on both $h_{i}$ and $h_{j}$ one says that the studied
network has two-point correlations \cite{newPRE2003a,bogPRE2003}.
To characterize this type of correlations one usually takes
advantage of the joint, two-dimensional distribution $R(h_i,h_j)$
that describes the probability of randomly chosen edge to connect
vertices labeled as $h_i$ and $h_j$.

\par In order to characterize network in a more detailed way the
concept of higher order correlations, given by multidimensional
probability distributions, should be used. In this paper we limit
ourselves to two-point correlations. The lack of higher order
correlations is ensured by the factorization of the conditional
distribution $R(h^{(1)}, h^{(2)},\dots,h^{(x)}/h)= R(h^{(1)}/h)
R(h^{(2)}/h) \dots R(h^{(x)}/h)$, that describes the probability
of a node of property $h$ to have $x$ neighbors labeled as
$h^{(1)}, h^{(2)},\dots,h^{(x)}$. Such networks, with the only
two-point correlations are called {\it Markovian}, due to the
reason that they are completely defined by the joint distribution
$R(h_{i},h_{j})$, which on its own turn completely determines the
conditional distribution $R(h_{j}/h_{i})$ and the distribution of
hidden variables $R(h)$. Relationships between $R(h_{i},h_{j})$,
$R(h_{j}/h_{i})$, $r_{ij}$ and $R(h)$ will be analyzed later.

\par After this short introduction to the concept of hidden variables
and to the problem of network correlations we may go back to the
main topic of the paper i.e. the interplay between different
levels of the network complexity. The approach has been initiated
by S\"oderberg \cite{sodPRE2002} and developed by Bogu\~{n}\'{a}
and Pastor-Satorras \cite{bogPRE2003}. Bogu\~{n}\'{a} and
Pastor-Satorras have concentrated on the question: how
correlations at the level of hidden variables affect pattern of
connections at the level of node degrees. Given $R(h_{i},h_{j})$
the authors have derived an analytical expression for the joint
distribution of degrees of the nearest neighbors $P(k_{i},k_{j})$.
In this paper, we ask the reverse question: what kind of hidden
correlations $R(h_{i},h_{j})$ produces the given pattern of node
degree correlations $P(k_{i},k_{j})$.

\par As a matter of fact, since most of us are better acquainted with
node degree notation than with abstract hidden variables, the {\it
reverse} approach seems to be very interesting, at least from the
methodological point of view. It is already well-known that there
exist degree correlations in real networks \cite{newPRL2002}. On
the other hand, due to the lack of data nothing is known about
correlation at the hidden level, from which the observed network
structure arises. The paper represents a small step towards
understanding the phenomenon of self-organization in complex
networks beyond the predominant approach of the so-called evolving
networks \cite{BAScience,BAPhysA1999}.

\par The paper is organized as follows. In Sec. \ref{boguna} we
review general results on correlated random networks with hidden
variables \cite{bogPRE2003}. The salient issues concerning
two-point correlations are discussed in this section. Sec.
\ref{poisson} is devoted to theoretical aspects of our {\it
reverse} problem. Some remarks on uncorrelated networks and a
practical algorithm of generating random networks with two-point
degree correlations are also given in this section. Finally, in
Sec. \ref{sum} we draw our conclusions.


\section{Correlated networks with hidden variables}\label{boguna}

\subsection{Construction procedure}\label{constr}

\par Probably the most attractive feature of random networks with
hidden variables is the construction procedure that consists of
only two steps:
\begin{itemize}
\item [i.] first, prepare $N$ nodes and assign them hidden
variables independently drawn from the probability distribution
$R(h)$,
\item [ii.] next, each pair of nodes $\{i,j\}$ link with a probability
$r_{ij}$.
\end{itemize}
Note that the above procedure does not include any ambiguous
instructions like {\it in the case of $\dots$ do $\dots$}. The
lack of ambiguities enables analytical treatment of the discussed
models.

\par The particularly simple model of sparse random networks with
hidden variables has been recently considered by Bogu\~{n}\'{a}
and Pastor-Satorras. Provided that the connection probability
scales according to
\begin{equation}\label{gij1}
r_{ij}=\frac{f(h_i,h_j)}{N},
\end{equation}
where $f(h_i,h_j)=f(h_j,h_i)$, the authors have shown that the
degree distribution of nodes characterized by a fixed value of
hidden variable $h$ is given by the Poisson distribution
\begin{equation}\label{prop}
g(k/h)=\frac{e^{-h}h^{k}}{k!}.
\end{equation}

\par The last result is very important because it joins the two
levels of the network complexity and allows to freely move between
the notation of hidden variables and the notation of node degrees
(see section \ref{poisson}). In particular, taking advantage of
(\ref{prop}), one can see that the average connectivity of nodes
characterized by $h$ is simply $h$
\begin{equation}\label{skh}
\langle k\rangle(h)=\sum_{k} k g(k/h)=h,
\end{equation}
i.e. hidden variables play the role of {\it desirable degrees}.
Next, due to (\ref{prop}), the degree distribution $P(k)$ gains a
simple form of convolution
\begin{equation}\label{Pk}
P(k)=\int R(h)g(k/h)dh.
\end{equation}
The last relation between both distributions $P(k)$ and $R(h)$
implies relation between their moments
\begin{equation}
\langle h^{n}\rangle=\langle k(k-1)\dots(k-n+1)\rangle,
\end{equation}
and respectively
\begin{equation}\label{mom}
\langle h\rangle=\langle k\rangle,\;\;\;\;\;\;\;\langle
h^{2}\rangle=\langle k(k-1)\rangle.
\end{equation}
As a matter of fact, in the case of other pair-related
distributions, like $R(h_i,h_j)$ and $P(k_i,k_j)$, it is also
possible to obtain relations similar to (\ref{Pk}). However,
before we proceed with the interplay between hidden variables and
node degrees, we would like to discuss the issue of two-point
correlations.

\subsection{Two-point correlations}

\par
\begin{figure} \epsfxsize=7.5cm \epsfbox{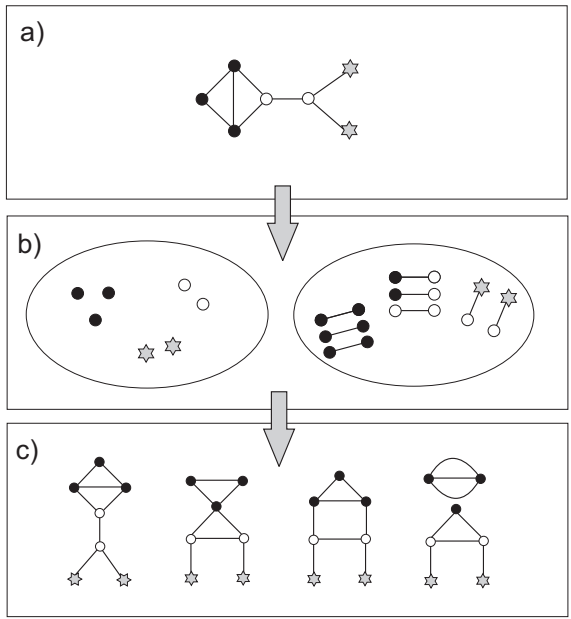}
\caption{Schematic representation of an ensemble of random
networks with a given two-point correlations at the level of
hidden variables. Detailed description of the figure is given in
the text.}\label{figexample}
\end{figure}

 One usually thinks about a network as about a collection of nodes
and links. Distribution of hidden variables $R(h)$ is the basic
characteristic of the set of nodes, whereas the joint distribution
$R(h_{i},h_{j})$, describing two-point correlations, applies to
the set of edges (see Fig.\ref{figexample}). Both distributions
express the probability that a random representative of its own
ensemble has assigned a given attribute i.e. property $h$ in the
case of node and pair of hidden variables $\{h_{i}$, $h_{j}\}$ in
the case of edge.

\par At the moment, let us briefly comment on the Fig.\ref{figexample}. The
second stage of the figure shows the set of nodes and the set of
links, both corresponding to the simple network presented at the
first stage of the same figure. Obviously, basing only on the two
sets is almost impossible to recreate the original network. Such a
representation related to a single network neglects much
information. What is lacking are higher order correlations. On the
other hand, the joint distribution $R(h_i,h_j)$ characterizing the
set of edges conveys all the information that is required to
construct an {\it ensemble} of Markovian random networks (see the
third stage of the Fig.\ref{figexample}). In fact, all the
calculations performed in this paper are related to such an
ensemble.

\par For a given network in the thermodynamic limit, both distributions
$R(h)$ and $R(h_i,h_j)$ may be defined in the following way
\begin{equation}\label{Gh1}
R(h)=\frac{N(h)}{N},
\end{equation}
and
\begin{equation}\label{Ghihj1}
R(h_i,h_j)=\frac{E(h_i,h_j)}{E},
\end{equation}
where $N(h)$ gives the number of nodes assigned by $h$ and $N=\int
N(h)dh$, whereas $E(h_i,h_j)$ describes the number of links
labeled as $\{h_{i},h_{j}\}$ and due to the normalization
condition $E=\int\int E(h_i,h_j)dh_{i}dh_{j}$.

\par The connection probability $r_{ij}$ (\ref{gij1}) may be simply
expressed by the ratio between the actual number of edges
connecting vertices of tags $h_i$ and $h_j$ and the maximum value
of this quantity
\begin{eqnarray}\label{gij2}
r_{ij}=\frac{E(h_i,h_j)}{E_{max}(h_i,h_j)}.
\end{eqnarray}
Since, during the network construction process one analyzes all
pairs of nodes, the maximum number of the considered connections
is given by $E_{max}=N(h_i)N(h_j)$. Now, taking advantage of
(\ref{Ghihj1}) the expression for $r_{ij}$ may by rewritten in the
following form
\begin{eqnarray}\label{gij3}
r_{ij}=\frac{\langle h\rangle R(h_i,h_j)}{N R(h_i)R(h_j)}.
\end{eqnarray}

\par To fulfill discussion on two point correlations it is necessary
to find relation between the two distributions (\ref{Gh1}) and
(\ref{Ghihj1}). The relation is encoded in the so-called detailed
balance condition
\begin{equation}\label{Ghcond2}
R(h_{j}/h_{i})=\frac{\langle h\rangle
R(h_{i},h_{j})}{h_{i}R(h_{i})}.
\end{equation}
Summing both sides of the last identity over $h_j$ ($\int
R(h_j/h_i)dh_j=1$) one obtains
\begin{equation}\label{Gh2}
R(h_{i})=\frac{\langle h\rangle}{h_{i}} \int R(h_{i},h_{j})dh_j.
\end{equation}

\par As a truth, almost all the above derivations that have been done for
hidden variables also hold for node degrees (see
\cite{bogEPJ2004}). In particular, the {\it degree} detailed
balance condition possesses the same form as (\ref{Ghcond2})
\begin{equation}\label{Pkcond1}
P(k_{j}/k_{i})=\frac{\langle k\rangle
P(k_{i},k_{j})}{k_{i}P(k_{i})}
\end{equation}
and respectively the degree distribution in Markovian networks is
given by the relation analogous to (\ref{Gh2})
\begin{equation}\label{Pk1}
P(k_{i})=\frac{\langle k\rangle}{k_{i}} \sum_{k_j} P(k_{i},k_{j}).
\end{equation}

\subsection{Interplay between hidden variables and node degrees}

\par Each node of the considered networks is characterized by two
parameters: hidden variable $h$ and degree $k$. The probability
$p(k\cap h)$ that two given parameters $h$ and $k$ meet together
in a certain node is described by the below identity
\begin{equation}\label{prop*}
p(k\cap h)=R(h)g(k/h)=P(k)g^{*}(h/k).
\end{equation}
The meaning of both conditional distributions $g(k/h)$ and
$g^{*}(h/k)$ is simple. The first distribution $g(k/h)$
(\ref{prop}) has just been introduced. It describes the
probability that a node labeled by $h$ has $k$ nearest neighbors.
The second distribution $g^{*}(h/k)$ is complementary to the first
one and gives the probability that a node with $k$ nearest
neighbors is labeled by $h$.

\par The knowledge of $g(k/h)$ and $g^{*}(h/k)$ allows one to find
the relation between the joint distributions $R(h_i,h_j)$ and
$P(k_{i},k_{j})$ characterizing pair correlations respectively at
the level of hidden variables and at the level of node degrees.
Simple arguments let one describe the interplay by the following
relation \cite{info}
\begin{equation}\label{joink1}
P(k_i,k_j)=\int\int g(k_i-1/h_i)R(h_i,h_j)g(k_j-1/h_j)dh_idh_j.
\end{equation}

The last expression states that if one knows hidden correlations
then it is possible to calculate degree correlations. Relating the
problem to the present {\it state of the art} in the field of
complex networks it is a very artificial situation. The concept of
networks with hidden variables is still little-known because most
of researchers are strongly attached to the node degree notation.
From this point of view, the reverse problem that would answer the
question: what kind of hidden correlations produces given degree
correlations, seems to be very attractive. In addition, since one
knows how to construct Markovian networks with hidden variables,
the solution of the above problem would also allow, in a simple
way, to generate networks with given two-point degree
correlations. We work out the problem in the next section entitled
"Degree correlations from hidden correlations".

In fact, the idea of generating networks with given degree
correlations by means of networks with hidden variables originates
from Bogu\~{n}\'{a} and Pastor-Satorras. The authors have argued
that since the conditional probability $g(h/k)$ is Poissonian
(\ref{prop}), the joint distribution for node degrees
(\ref{joink1}) approaches the joint distribution for hidden
variables for $k_i,k_j\gg 1$
\begin{equation}\label{pom2}
P(k_i,k_j)\simeq R(k_i,k_j),
\end{equation}
and respectively asymptotic behavior of the degree distribution is
given by
\begin{equation}\label{pom3}
P(k_i)\simeq R(k_i).
\end{equation}
The range of convergence of the two distributions has been
estimated by Dorogovtsev \cite{dorCM2003}, who has shown that both
approximations (\ref{pom2}) and (\ref{pom3}) are only acceptable
when $R(h_i,h_j)$ and $R(h)$ are sufficiently slowly decreasing.

\section{Hidden correlations from degree correlations}\label{poisson}

\subsection{Depoissonization}\label{depo}

At the moment, let us examine Eq. (\ref{Pk}) more carefully.
Depending on whether $h$ is discrete or continuous variable the
expression for $P(k)$ is respectively discrete or integral
transform with Poissonian kernel \cite{PT1}. The issue of
determining $R(h)$ from $P(k)$ is simply the problem of finding
the inverse transform \cite{PT2} and one can show that for a given
$P(k)$ there exists a unique $R(h)$ which satisfies (\ref{Pk}).
When $h$ is continuous one gets (see {\it Appendix})
\begin{equation}\label{Rh1i}
R(h)=e^{h}\mathcal{F}^{-1}[G(ix)],
\end{equation}
where $\mathcal{F}^{-1}$ denotes the inverse Fourier transform and
$G(x)$ describes generating function for the degree distribution
$P(k)$
\begin{equation}\label{G0}
G(x)=\sum_{k}x^{k}P(k).
\end{equation}
In the case of discrete $h$, the inverse Poisson transform is
given by the formula
\begin{equation}\label{Rh1d}
R(h)=e^{h}\mathcal{Z}^{-1}[G(-\ln x)],
\end{equation}
where $\mathcal{Z}^{-1}$ describes the inverse Z-transform.

The same applies for the joint degree distributions characterizing
two-point correlations. Since
\begin{equation}
g(k-1/h)=\frac{k}{h}\;g(k/h),
\end{equation}
the formula (\ref{joink1}) may be rewritten as
\begin{equation}\label{joink2}
\frac{P(k_i,k_j)}{k_ik_j}=\int\int
g(k_j/h_j)\frac{R(h_i,h_j)}{h_ih_j}g(k_i/h_i)dh_idh_j.
\end{equation}
Now, it is easy to see that $P(k_i,k_j)$ is connected with
$R(h_i,h_j)$ by the two-dimensional Poisson transform and
\begin{equation}\label{Rhihjdepo}
\frac{R(h_i,h_j)}{h_ih_j}=e^{h_i+h_j}\;\mathcal{F}^{-1}[\mathcal{F}^{-1}[G^{*}(ix,iy)]],
\end{equation}
where
\begin{equation}\label{G*}
G^{*}(x,y)=\sum_{k_i}\sum_{k_j}x^{k_i}y^{k_j}\frac{P(k_i,k_j)}{k_ik_j}.
\end{equation}
The last two expressions can be simply translated for the case of
discrete hidden variables. Since however Fourier transforms are
more convenient to work with then Z-transforms, we have decided to
pass over such a reformulation.

\subsection{Some remarks on uncorrelated networks}\label{uncor}

As we said at the beginning of this paper - the lack of pair
correlations at the level of hidden variables means that the
conditional probability $R_{o}(h_j/h_i)$ does not depend on $h_i$
(to differentiate between uncorrelated and correlated networks the
characteristics related to the former case have been denoted by
the subscript 'o'). In fact, it is simple to show that
\begin{equation}\label{Rocond}
R_{o}(h_j/h_i)=\frac{h_j}{\langle h\rangle}R_{o}(h_j),
\end{equation}
and respectively the joint distribution (\ref{Ghcond2}) gains a
factorized form
\begin{equation}\label{Rojoint}
R_{o}(h_i,h_j)=\frac{h_{i}R_{o}(h_{i})h_{j}R_{o}(h_{j})}{\langle
h\rangle^{2}}.
\end{equation}
Inserting the above expression into (\ref{gij2}) one gets the
formula for the connection probability in uncorrelated networks
\begin{equation}\label{gijo}
r^{o}_{ij}(h_i,h_j)=\frac{h_{i}h_{j}}{\langle h\rangle N}.
\end{equation}

Now, the question is: does the lack of pair correlations at the
hidden level translates to the lack of pair correlations between
degrees of the nearest neighbors? Inserting (\ref{Rojoint}) into
(\ref{joink1}), and then taking advantage of the degree detailed
balance condition (\ref{Pkcond1}) one gets the answer
\begin{equation}\label{Pocond}
P_{o}(k_j/k_i)=\frac{k_j}{\langle k\rangle}P_{o}(k_j),
\end{equation}
i.e. the lack of hidden correlations results in the lack of node
degree correlations. Thus, in order to generate uncorrelated
network with a given degree distribution $P(k)$ one has to:
\begin{itemize}
\item[i.] prepare the desired number of nodes $N$,
\item[ii.] label each node by a hidden variable randomly taken from the
distribution $R(h)$ given by (\ref{Rh1i}) or (\ref{Rh1d}),
\item[iii.] each pair of nodes link with the probability
$r^{o}_{ij}$ (\ref{gijo}).
\end{itemize}
(See also other methods for generating random uncorrelated
networks with a given degree distribution \cite{newPRE2001,
krzPRE2001}).

Note also that, due to properties of the Poisson propagator
(\ref{skh})
\begin{equation}
h_{i}=\langle k(h_i)\rangle\;\;\;\;\mbox{and}\;\;\;\;
h_{j}=\langle k(h_j)\rangle,
\end{equation}
therefore the connection probability $r_{ij}^{o}(h_i,h_j)$
(\ref{gijo}) may be considered as the connection probability
$p_{ij}^{o}(k_i,k_j)$ between two nodes of degrees $k_i$ and $k_j$
averaged over all pairs of nodes possessing hidden variables
respectively equal to $h_i$ and $h_j$
\begin{equation}\label{gijo1}
r^{o}_{ij}(h_i,h_j)=\sum_{k_i} \sum_{k_j}
g(k_i/h_i)\;p^{o}_{ij}(k_i,k_j)\;g(k_j/h_j),
\end{equation}
where
\begin{equation}\label{pijo1}
p^{o}_{ij}(k_i,k_j)=\frac{k_ik_j}{\langle k\rangle N}.
\end{equation}
The last expression describing connection probability in
uncorrelated sparse networks has been already used by several
authors (in particular see
\cite{afPRE2004,afCM2002,chung2002,newPRE2003b}).

\subsection{Examples of uncorrelated networks}

\par{\it 1. Classical random graphs of Erd\"{o}s and R\`{e}nyi.} The
degree distribution in ER model is Poissonian
\begin{equation}
P(k)=\frac{e^{-\langle k\rangle}\langle
k\rangle^{k}}{k!}\;\;\;\;\;\;\; k\geq 0.
\end{equation}
The first step towards calculation of the required distribution
$R(h)$ is finding characteristic function for Poisson
distribution. Taking advantage of (\ref{G0}) one gets
\begin{equation}
G(x)=e^{\langle k\rangle(x-1)}.
\end{equation}
Now, inserting the exponential function into (\ref{Rh1i}) or
(\ref{Rh1d}) one can see that the distribution of hidden variables
in classical random graphs is respectively given by the Dirac's
delta function (in the case of continuous $h$) or the Kronecker
delta (in the case of discrete $h$)
\begin{equation}
R(h)=\delta(h-\langle k\rangle).
\end{equation}
The above result means that all vertices are equivalent and the
connection probability at the level of hidden variables
(\ref{gijo}) is given by
\begin{equation}
r^{o}_{ij}=\frac{\langle k\rangle}{N}.
\end{equation}

\begin{figure} \epsfxsize=7.8cm \epsfbox{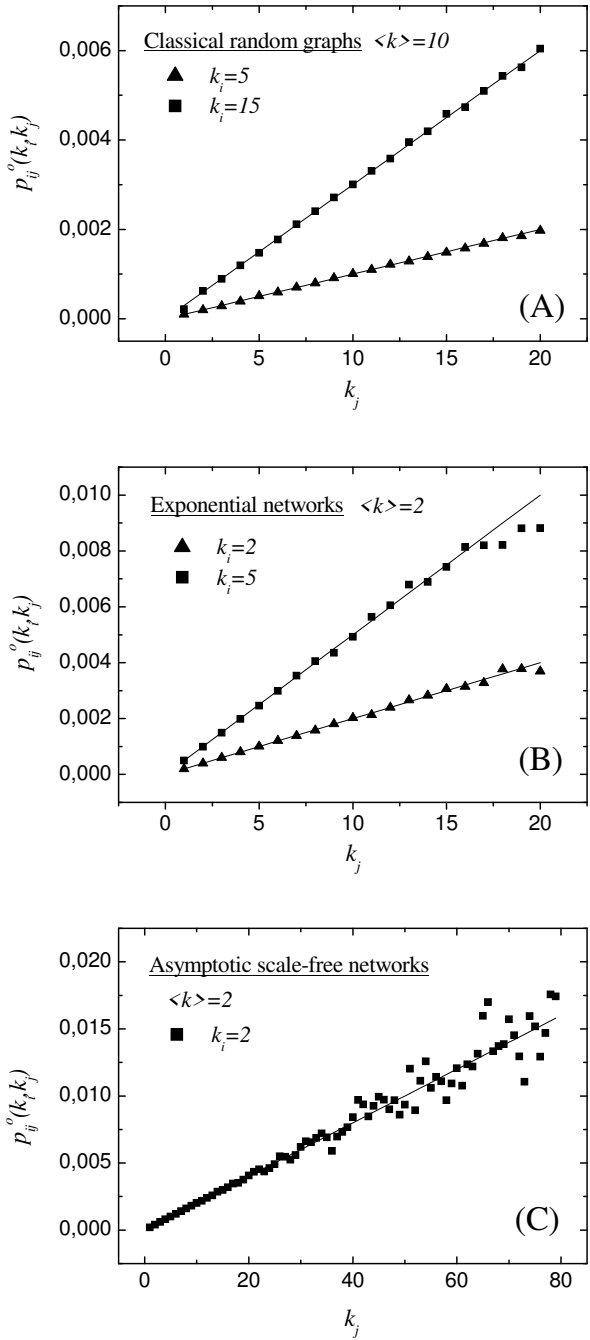}
\caption{Connection probability $p_{ij}^{o}(k_i,k_j)$ in classical
random graphs (A), uncorrelated networks with exponential degree
distribution (\ref{PkEXP}) (B) and asymptotic scale-free networks
(\ref{Pksfa1}) (C). Points represent results of numerical
calculations averaged over $500$ networks of size $N=5000$.}
\label{figALLpij}
\end{figure}

Fig. \ref{figALLpij} presents probability $p^{o}_{ij}(k_i,k_j)$ of
a connection between two nodes characterized by degrees $k_i$ and
$k_j$ in ER model and other uncorrelated networks. As one can see,
there exists a very good agreement between the formula
(\ref{pijo1}) and results of numerical calculation.

\par
\begin{figure} \epsfxsize=7.8cm \epsfbox{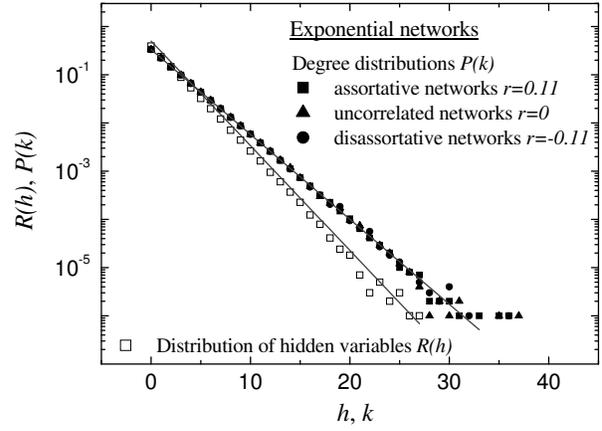}
\caption{Distribution of hidden variables and degree distributions
obtained from numerical simulations of uncorrelated and correlated
exponential networks.}\label{figEXPPk}
\end{figure}

{\it 2. Networks with $P(k)$ given by exponential
distribution}. Now, let us suppose that
\begin{equation}\label{PkEXP}
P(k)=\frac{\langle k\rangle^{k}}{(1+\langle
k\rangle)^{k+1}}\;\;\;\;\;\;\; k\geq 0.
\end{equation}
The generating function of the above degree distribution is given
by the sum of the geometric series
\begin{equation}
G(x)=\frac{1}{1+\langle k\rangle-\langle k\rangle x}
\end{equation}
and (\ref{Rh1i}) provides the exponential distribution of hidden
variables (see Fig. \ref{figEXPPk})
\begin{equation}\label{RhEXP}
R(h)=\frac{e^{-h/\langle k\rangle}}{\langle
k\rangle},\;\;\;\;\;\;\; h\geq 0.
\end{equation}

\par{\it 3. Scale-free networks.\label{sectionSF}} In mathematical terms, the
scale-free property translates into a power-law degree
distribution
\begin{equation}\label{Psf}
P(k)=\frac{A}{k^\alpha},\;\;\;\;\;\;\; k\geq 1
\end{equation}
where $\alpha$ is a characteristic exponent and $A$ represents
normalization constant. Generating function for this distribution
is given by the polylogarithm
\begin{equation}\label{Gsf}
G(x)=A\sum_{k=1}^{\infty}\frac{x^k}{k^\alpha}=A\;{\mbox L}{\mbox
i}_\alpha(x).
\end{equation}
To derive distribution of hidden variables that leads to
uncorrelated scale-free network one has to find the inverse
Fourier transform of the polylogarithm with imaginary argument
\begin{equation}\label{Rhsf}
R(h)=e^{h}\mathcal{F}^{-1}[{\mbox L}{\mbox i}_\gamma(ix)],
\end{equation}
or the adequate inverse Z-transform.

\par
\begin{figure} \epsfxsize=7.8cm \epsfbox{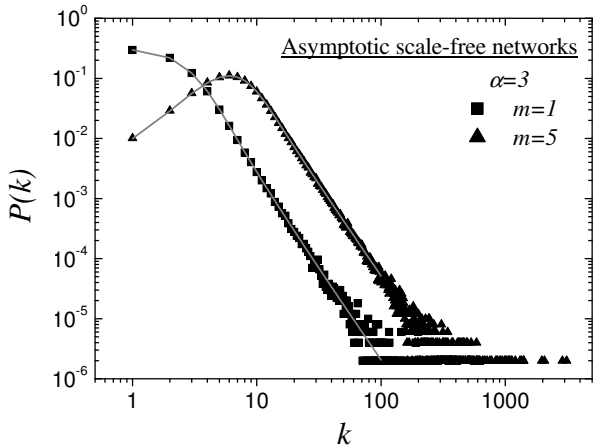}
\caption{Degree distributions in uncorrelated networks with
scale-free distributions of hidden variables (\ref{Rhsfa1}).
Points represent results of numerical calculations whereas solid
lines correspond to Eq. (\ref{Pksfa1}).} \label{figPkSF}
\end{figure}

 Unfortunately, it does not appear that a closed-form solutions
for both inverse transforms can be simply obtained. Nevertheless,
some asymptotic results for scale-free networks can be derived. In
particular, one can show that power law distribution of hidden
variables
\begin{equation}\label{Rhsfa1}
R(h)=\frac{(\alpha-1)m^{(\alpha-1)}}{h^\alpha},\;\;\;\;\;\;\;
h\geq m,
\end{equation}
leads to asymptotic scale-free networks with degree distribution
given by
\begin{equation}\label{Pksfa1}
P(k)=(\alpha-1)m^{(\alpha-1)}\frac{\Gamma(k-\alpha+1,m)}{k!},
\end{equation}
where $\Gamma(x,y)$ is incomplete gamma function. In the limit of
large degrees $k\gg 1$ the above degree distribution decays as
$P(k)\sim k^{-\alpha}$ (see Fig. \ref{figPkSF}). The effect of
structural cutoffs in power law distributions of hidden variables
with $\alpha<3$ (\ref{Rhsfa1}), imposing the largest hidden
variable to scale as $h_{max}\sim \sqrt{N}$ (the relation follows
from $r_{ij}^{o}\leq 1$ (\ref{gijo})) \cite{bogEPJ2004,catPRE2004,
burdaPRE2003}, does not represent any problem in the studied
formalism. The effect of $h_{max}$ in the scale-free $R(h)$ may be
considered as an {\it exponential} cut-off
\begin{equation}\label{Rhsfa2}
R(h)=\frac{(\alpha-1)m^{(\alpha-1)}}{h^\alpha}\exp\left(-\frac{h}{h_{max}}\right).
\end{equation}
Due to properties of the Poisson transform \cite{PT1}, the above
$R(h)$ results in truncated degree distribution (\ref{Pksfa1})
\begin{equation}\label{Pksfa2}
P(k)=\frac{\Gamma(k-\alpha+1,m)}{k!}
\frac{(\alpha-1)m^{(\alpha-1)}}{(1/h_{max}+1)^{k-\alpha+1}}.
\end{equation}
As expected (see discussion of Eq. (\ref{pom3})), in the limit of
large degrees the last formula approaches (\ref{Rhsfa2}).

\par Before we finish with scale-free networks let us, once again,
concentrate on the Fig. \ref{figPkSF}. The figure presents degree
distributions $P(k)$ in sparse networks with scale-free
distributions of hidden attributes $R(h)$ (\ref{Rhsfa1}). One can
observe that although both distributions converge in the limit of
large degrees, there exist serious deviations between
(\ref{Rhsfa1}) and (\ref{Pksfa1}) in the limit of small degrees.
The relative behavior of the two distributions let one expect that
the correct $R(h)$ reproducing the {\it pure} scale-free degree
distribution $P(k)$ should describe a kind of {\it condensate}
with a huge number of nodes characterized by a very low values of
hidden attributes. On the other hand, despite the {\it ambiguous}
behavior of $P(k)$ for small degrees, the power-law tail is
interesting on his own right. A number of real networks have fat
tailed degree distributions. The above allows us to deduce on fat
tailed distributions of underlying hidden attributes assigned to
individuals co-creating the considered systems.

\subsection{How to generate correlated networks with a given degree
correlations\label{corr}}

Here, we again make use of the depoissonization idea proposed in
Sec. \ref{depo}. The procedure of generating random networks with
a given two-point, degree correlations $P(k_i,k_j)$ is as follows:
\begin{itemize}
\item[i.] first, prepare $N$ nodes, \item[ii.] next, label each
node by a hidden variable randomly taken from the distribution
$R(h)$ (\ref{Rh1i}), where $P(k)$ is is given by (\ref{Pk1}),
\item[iii.] finally, link each pair of nodes with the probability
$r_{ij}$ (\ref{gij2}), where $R(h_i,h_j)$ is calculated according
to the formula (\ref{Rhihjdepo}).
\end{itemize}

Although very clear, the above procedure suffers a certain
inconvenience: given a joint degree distribution $P(k_i,k_j)$ the
closed form solution for $R(h_i,h_j)$ (\ref{Rhihjdepo}) is often
hard to get. Since however there exists a number of algorithms for
numerical inversion of Fourier transform the above does not
represent a real problem.

\subsection{Examples of correlated networks\label{correx}}

To make derivations of previous sections more concrete, we should
immediately introduce some examples of correlated networks. In
order to simplify the task we will take advantage of general
patterns for joint degree distributions with two-point assortative
(a) and disassortative (d) correlations that was proposed by
Newman \cite{newPRE2003a}
\begin{eqnarray}\label{Pkd0}
P^{d}(k_i,k_j)=\widetilde{P}(k_i)\widetilde{Q}(k_j)+
\widetilde{Q}(k_i)\widetilde{P}(k_j)-
\widetilde{Q}(k_i)\widetilde{Q}(k_j)\;\;\;
\end{eqnarray}
and
\begin{eqnarray}\label{Pka0}
P^{a}(k_i,k_j)=2\widetilde{P}(k_i)\widetilde{P}(k_j)-P^{d}(k_i,k_j),
\end{eqnarray}
where $\widetilde{P}(k)$ and $\widetilde{Q}(k)$ are arbitrary
distributions such that $\sum_k \widetilde{P}(k)=\sum_k
\widetilde{Q}(k)=1$. Assortativity / correlation coefficients
\cite{newPRL2002} for the above distributions are respectively
equal to
\begin{equation}\label{defr}
r^d=-\frac{(\mu_p-\mu_q)^2}{\sigma_p^2}\;\;\;\;\;\;\;\;
\mbox{and}\;\;\;\;\;\;\;\;r^a=-r^d,
\end{equation}
where $\mu_p$ represents the expectation value for
$\widetilde{P}(k)$, $\mu_q$ has the same meaning for
$\widetilde{Q}(k)$, whereas $\sigma_p^2$ corresponds to the
variance of $\widetilde{P}(k)$. Now, in order to facilitate
further calculations let us assume that
\begin{equation}\label{pomc1}
\widetilde{P}(k)=\frac{k}{\langle
k\rangle}P(k)\;\;\;\;\;\;\mbox{and}
\;\;\;\;\;\;\widetilde{Q}(k)=\frac{k}{\langle q\rangle}Q(k),
\end{equation}
where
\begin{equation}
\langle k\rangle=\sum_k kP(k)\;\;\;\;\;\;\;\mbox{and}
\;\;\;\;\;\;\;\langle q\rangle=\sum_k kQ(k).
\end{equation}
Note that there is no conflict of notation in the last assignment.
Putting the two expressions (\ref{pomc1}) into $P^a$ or $P^d$ and
then taking advantage of the degree detailed balance condition
(\ref{Pk1}) one can easily check that $P(k)$ corresponds to degree
distribution. Now, given $P(k)$ one can execute the first two
steps (i. and ii.) of the construction procedure described in the
previous section.

Inserting the relations (\ref{pomc1}) into (\ref{Pkd0}) and
(\ref{Pka0}) one obtains (compare it with (\ref{joink2}))
\begin{eqnarray}\label{Pkd1}
\frac{P^{d}(k_i,k_j)}{k_ik_j}=\frac{P(k_i)Q(k_j)}{\langle
k\rangle\langle q\rangle}+\frac{Q(k_i)P(k_j)}{\langle
k\rangle\langle q\rangle}-\frac{Q(k_i)Q(k_j)}{\langle q\rangle^2}
\end{eqnarray}
and
\begin{eqnarray}\label{Pka1}
\frac{P^{a}(k_i,k_j)}{k_ik_j}=2\frac{P(k_i)P(k_j)}{\langle
k\rangle^2}-\frac{P^{d}(k_i,k_j)}{k_ik_j}.
\end{eqnarray}
Due to linearity of the Poisson transform the joint distributions
(\ref{Pkd1}) and (\ref{Pka1}) turn out to be particularly useful
for our purposes. The usefulness means that whenever closed form
solutions for the inverse Poisson transforms of $P(k)$ and $Q(k)$
exist, one can also obtain closed form solutions for the joint
hidden distributions $R^d(h_i,h_j)$ and $R^a(h_i,h_j)$
(\ref{Rhihjdepo}).

Generating functions $G^{*}(x,y)$ (\ref{G*}) for (\ref{Pkd1}) and
(\ref{Pka1}) are respectively given by
\begin{eqnarray}\label{G*Pkd1}
G^{*}_d(x,y)=\frac{G_p(x)G_q(y)}{\langle k\rangle\langle
q\rangle}+\frac{G_q(x)G_p(y)}{\langle k\rangle\langle
q\rangle}-\frac{G_q(x)G_q(y)}{\langle q\rangle^2}
\end{eqnarray}
and
\begin{eqnarray}\label{G*Pka1}
G^{*}_a(x,y)=2\frac{G_p(x)G_p(y)}{\langle
k\rangle^2}-G^{*}_d(x,y),
\end{eqnarray}
where
\begin{eqnarray}\label{Gpq}
G_p(x)=\sum_k x^kP(k)\;\;\;\;\mbox{and}\;\;\;\;G_q(x)=\sum_k
x^kQ(k).
\end{eqnarray}
Although visually quite complicated all the above formulas are in
fact very simple. Now, in order to perform the last step (iii.) of
our procedure aiming at constructing correlated networks, one has
to calculate the joint distribution $R(h_i,h_j)$. Taking advantage
of (\ref{Rhihjdepo}) one gets
\begin{equation}\label{Rhd1}
R^{d}(h_i,h_j)=\widetilde{R}(h_i)\widetilde{S}(h_j)+
\widetilde{S}(h_i)\widetilde{R}(h_j)-
\widetilde{S}(h_i)\widetilde{S}(h_j)
\end{equation}
and
\begin{equation}\label{Rha1}
R^{a}(h_i,h_j)=2\widetilde{R}(h_i)\widetilde{R}(h_j)-R^{d}(h_i,h_j),
\end{equation}
where, similarly to (\ref{pomc1}) one has
\begin{equation}\label{pomc2}
\widetilde{R}(h)=\frac{h}{\langle k\rangle}R(h)\;\;\;
\mbox{and}\;\;\;R(h)=e^{h}\mathcal{F}^{-1}[G_p(ix)],
\end{equation}
and also
\begin{equation}\label{pomc3}
\widetilde{S}(h)=\frac{h}{\langle q\rangle}S(h)\;\;\;
\mbox{and}\;\;\;S(h)=e^{h}\mathcal{F}^{-1}[G_q(ix)].
\end{equation}
Note, that $R(h)$ given in (\ref{pomc2}) expresses distribution of
hidden variables in the considered correlated networks (i.e. the
inverse Poisson transform of the degree distribution $P(k)$).

\begin{figure} \epsfxsize=7.8cm \epsfbox{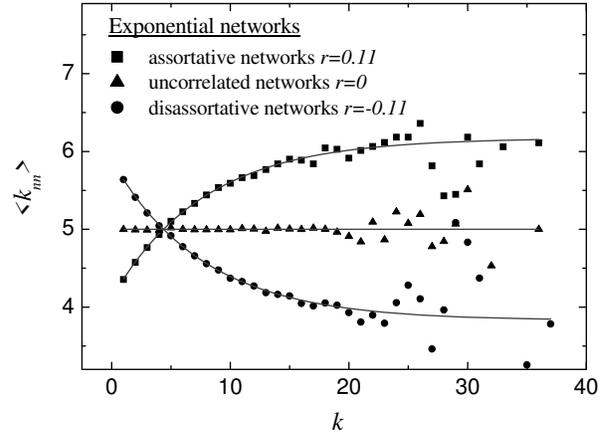}
\caption{Average nearest neighbor degree $\langle
k_{nn}\rangle(k)$ for assortative, uncorrelated and disassortative
networks with exponential degree distributions (see section
\ref{correx})}\label{figEXPknn}
\end{figure}

Now, let us translate the general considerations into a specific
example. Suppose that we are interested in networks with
exponential degree distribution (\ref{PkEXP})
\begin{equation}\label{PkEXP1}
P(k)=\frac{\langle k\rangle^{k}}{(1+\langle
k\rangle)^{k+1}}\;\;\;\;\;\;\; k\geq 0.
\end{equation}
Distribution of hidden variables in such networks is given by
(\ref{RhEXP})
\begin{equation}\label{RhEXP1}
R(h)=\frac{e^{-h/\langle k\rangle}}{\langle
k\rangle},\;\;\;\;\;\;\; h\geq 0.
\end{equation}
For mathematical simplicity, let us assume that the distribution
$Q(k)$ responsible for correlations is also exponential
\begin{equation}\label{QkEXP1}
Q(k)=\frac{\langle q\rangle^{k}}{(1+\langle
q\rangle)^{k+1}}\;\;\;\;\;\;\; k\geq 0.
\end{equation}
Given $P(k)$ and $Q(k)$ one has to ensure that the joint hidden
distributions (\ref{Rhd1}) and (\ref{Rha1}) and also the
connection probability $r_{ij}$ (\ref{gij2}) are positive and
smaller than $1$. It is easy to check, that in our case the
condition translates into the relation
\begin{equation}\label{xx}
\frac{1}{\sqrt{2}}\leq\frac{\langle q\rangle}{\langle
k\rangle}\leq 1.
\end{equation}

Now, let us briefly examine the role of $Q(k)$. Depending on the
value of $\langle q\rangle$ one obtains stronger or weaker
correlations characterized by (\ref{defr})
\begin{equation}\label{rour}
r^d=-\frac{2(\langle k\rangle-\langle q\rangle)^2} {\langle k
\rangle(\langle k\rangle+1)}\;\;\;\;\;\;\mbox{and}\;\;\;\;\;\;
r^a=-r^d.
\end{equation}
In particular, given $\langle k\rangle$ the expression (\ref{xx})
provides possible values for the correlation coefficients that may
be reproduced in the considered networks
\begin{equation}\label{rexample}
0\leq r\leq \frac{\langle k\rangle}{\langle k\rangle+1}
(\sqrt{2}-1)^2,
\end{equation}
where $r=|r^d|=|r^a|$.

In order to check the validity of our derivations we have
performed numerical simulations of correlated networks
(\ref{Pkd0}) and (\ref{Pka0}) with partial distributions given by
(\ref{PkEXP1}) and (\ref{QkEXP1}). Simulations were done for
networks ( both assortative and disassortative) of size $N=10^4$,
$\langle k\rangle=2$ and $\langle q\rangle=\sqrt{2}$. Given
$\langle k\rangle=2$, the value of $\langle q\rangle=\sqrt{2}$
corresponds to the maximum value of the correlation coefficient
$|r_{max}|=0.11$ (\ref{rexample}). At Fig. \ref{figEXPPk} we
depict results corresponding to degree distributions
(\ref{PkEXP1}) in the considered networks. Fig. \ref{figEXPknn}
presents effects of node degree correlations expressed by the
average degree of the nearest neighbor (ANND) that is defined as
\begin{equation}\label{knndef}
\langle k_{nn}\rangle(k)=\sum_{k_i}k_iP(k_i/k).
\end{equation}
It is already well-known that in the case of uncorrelated networks
(\ref{Pocond}) the ANND does not depend on $k$
\begin{equation}
\langle k_{nn}^o\rangle(k)=\frac{\langle k^2\rangle}{\langle
k\rangle},
\end{equation}
whereas in the case of assortatively (disassortatively) correlated
systems it is an increasing (decreasing) function of $k$. One can
find that in our example the corresponding functions are given by
the below formulas
\begin{eqnarray}\label{knno}
\langle k_{nn}^o\rangle(k)&=&1+2\langle k\rangle,\\
\langle k_{nn}^d\rangle(k)&=&\langle k_{nn}^o\rangle(k)-f(k),\\
\langle k_{nn}^a\rangle(k)&=&\langle k_{nn}^o\rangle(k)+f(k),
\end{eqnarray}
where
\begin{equation}
f(k)=2(\langle k\rangle-\langle q\rangle)\left[1-
\left(\frac{\langle q\rangle}{\langle k\rangle}\right)^{k-1}
\left(\frac{1+\langle k\rangle}{1+\langle
q\rangle}\right)^{k+1}\right].
\end{equation}
As one can see (Fig. \ref{figEXPknn}), the fit between computer
simulations and the above analytical expressions is very good,
certifying the validity of the proposed algorithm (section
\ref{corr}) for generating random networks with a given degree
correlations.

\section{Conclusions}\label{sum}

In this paper we refer to the set of articles devoted to the
so-called random networks with hidden variables. Importance of
this paper consists in ordering certain significant issues related
to both uncorrelated and correlated networks. In particular, we
show that networks being uncorrelated at the hidden level are also
lacking in correlations between node degrees. The observation
supported by the depoissonization idea (section \ref{depo}) allows
to extract distribution of hidden variables from a given node
degree distribution. Until now the distribution of hidden
variables required for generation of a given degree sequence had
to be guessed. From this point of view our findings complete the
algorithm for generating random uncorrelated networks that was
suggested by other authors \cite{sodPRE2002,chung2002}. We also
show that the connection probability in sparse uncorrelated
networks is factorized function of node degrees (\ref{pijo1}).

In this paper we also carefully analyze the interplay between
hidden attributes and node degrees. We show how to extract hidden
correlations from degree correlations and how to freely move
between the two levels of the networks complexity. Our derivations
provide mathematical background for the algorithm for generating
correlated networks that was proposed by Bogu\~{n}\'{a} and
Pastor-Satorras \cite{bogPRE2003}.

\section{Acknowledgments}

The authors thank Prof. Janusz A. Ho\l yst for useful discussions
and comments. A.F. acknowledges financial support from the
Foundation for Polish Science (FNP 2005) and the State Committee
for Scientific Research in Poland (Grant No. 1P03B04727). The work
of P.F. has been supported by European Commission Project CREEN
FP6-2003-NEST-Path-012864.

\section{Appendix}

The inverse Poisson transform for the case of continuous $h$
(\ref{Rh1i}) has been derived by E. Wolf and C.L. Mehta in 1964
\cite{PT2}. Below we outline the derivation, and then adopt it for
the case of discrete $h$ (\ref{Rh1d}).

Our aim it to inverse the formula
\begin{equation}
P(k)=\int_{0}^{\infty} \frac{e^{-h}h^{k}}{k!}R(h)dh.
\end{equation}
Let
\begin{eqnarray}
F(x)&=&\int_{0}^{\infty}e^{ixh}R(h)e^{-h}dh\nonumber\\&=&
\int_{0}^{\infty}\sum_{k=0}^{\infty}\frac{(ixh)^k}{k!}R(h)e^{-h}dh\\&=&
\sum_{k=0}^{\infty}(ix)^kP(k)=G(ix)\nonumber,
\end{eqnarray}
where $G$ (\ref{G0}) represents generating function for the degree
distribution $P(k)$. Then by the Fourier inversion formula one
gets the expression (\ref{Rh1i})
\begin{equation}
R(h)=e^{-h}\mathcal{F}^{-1}[G(ix)].
\end{equation}

The above derivation can be simply adopted for the case of
discrete transform with Poissonian kernel i.e.
\begin{equation}\label{dPT1}
P(k)=\sum_{h=0}^{\infty}\frac{e^{-h}h^k}{k!}R(h).
\end{equation}
Let us introduce an auxiliary sequence
\begin{equation}\label{dPT2}
J(h)=e^{-h}R(h).
\end{equation}
It is easy to show that the Z-transform of this sequence equals to
generating function of the degree distribution (\ref{G0})
\begin{eqnarray}
\mathcal{Z}[J(h)]&=&\sum_{h=0}^{\infty}\frac{J(h)}{s^h}\\&=&
\sum_{h=0}^{\infty}\left(\sum_{l=0}^{\infty}\frac{(-h\ln
s)^l}{l!}\right)J(h)\\&=&\sum_{l=0}^{\infty}(\ln s^{-1})^l
P(l)=G(\ln s^{-1}).
\end{eqnarray}
Applying the inverse Z-transform to the last expression one
obtains the formula (\ref{Rh1d}) describing distribution of hidden
variables
\begin{equation}\label{dPT4}
R(h)=e^h\mathcal{Z}^{-1}[G(\ln s^{-1})].
\end{equation}

\end{document}